
\input phyzzx.tex


\def\O{{\cal O}}

\def\Jpsi{{J/\psi}}

\def\half{{\textstyle {1\over 2}}}
\def\gev{{\rm GeV}}

\def\PL #1 #2 #3 {Phys.~Lett.~{\bf#1}\ (#2) #3}
\def\NP #1 #2 #3 {Nucl.~Phys.~{\bf#1}\ (#2) #3}
\def\PR #1 #2 #3 {Phys.~Rev.~{\bf#1}\ (#2) #3}
\def\PP #1 #2 #3 {Phys.~Rep.~{\bf#1}\ (#2) #3}
\def\PRL #1 #2 #3 {Phys.~Rev.~Lett.~{\bf#1}\ (#2) #3}
\def\ZP #1 #2 #3 {Z.~Phys.~{\bf#1}\ (#2) #3}

\def\refmark#1{$^{\,#1}$}

\hsize=6truein \vsize=8.5truein \voffset 3.2truecm \hoffset 3.7truecm

\overfullrule 0pt


\def\etal{{\it et. al.}}

\doublespace
\Pubnum{HU-TFT-93-50\cr
hep-ph/9310214}
\date{October 1993}

 \titlepage
\vfill
 \title{%
QCD EFFECTS IN PARTICLE PRODUCTION ON NUCLEI%
\foot{Invited talk at the Cracow Workshop on Multiparticle Production,
Cracow May 1993.}}

 \author{Paul Hoyer}
\address{Department of Physics \break 
University of Helsinki, Helsinki, Finland}
\vfill

{\advance\leftskip by .5in
\advance\rightskip by .5in
\tenpoint\baselineskip 12pt\smallskip
\centerline{ABSTRACT}

I discuss some questions related to hard scattering processes in nuclei
and corrections to the leading twist approximation. The QCD
factorization theorem requires that high energy partons do not lose energy
while traversing the nucleus. I explain the physical reason for this. The
theorem also states that spectator partons, not involved in the hard
collision, have no influence on the inclusive cross section. Important
spectator effects are, however, seen in the data for certain reactions and
in some kinematical regions. I discuss the reasons from a phenomenological
and theoretical point of view. Finally, I mention some methods for analyzing
hard processes in regions of very large $x$, where the leading twist terms
are not dominant.
\smallskip}

\REF\revs{S. J. Brodsky, Invited talk at Quark Matter '91, \NP A544 1992
223 ; L. Frankfurt and M. Strikman, Prog. Part. Nucl. Phys. {\bf 27} (1991)
135; L. Frankfurt, G. A. Miller and M. Strikman, Comm. Nucl. Part. Phys.
{\bf 21} (1992) 1; N. N. Nikolaev, Comm. Nucl. Part. Phys. {\bf 21} (1992)
41 and preprint LANDAU 9-93 (Feb. 1993); B. Z. Kopeliovich, Sov. J. Nucl.
Phys. {\bf 55} (1992) 752 and Torino preprint DFTT 24/93 (May 1993).}

\REF\fac{
J. C. Collins, D. E. Soper and G. Sterman, published in {\it Perturbative
QCD}, edited by A.H. Mueller, World Scientific, (1989);
G. Bodwin, Phys. Rev. {\bf D31} (1985) 2616 and {\bf D34}  (1986E) 3932;
J. Qiu and G. Sterman, Nucl. Phys. {\bf B353} (1991) 105, {\it ibid.}, 137.}

\baselineskip 14pt

\vskip 1cm
Studies of the dependence of hard collisions on the nuclear number of the
target and/or projectile have revealed a large number of interesting,
diverse and often surprising QCD effects. The nucleus has been, and will
continue to be, a versatile tool for uncovering reaction mechanisms and
indicating where our standard methods are inadequate. In this talk I shall
confine myself to two particular questions, related to the energy loss of
partons in nuclei and the importance of higher twist corrections due to
spectator parton interactions. For a wider view of nuclear effects, I
refer to recent reviews\refmark\revs.

\break 
{\bf 1. Parton Energy Loss in Nuclei}

At leading twist, \ie, up to corrections of $\O(1/Q^2)$ in the hard scale
$Q$, the dependence on hard and soft physics factorizes in a simple way in
QCD\refmark\fac. Thus, \eg, the cross-section for the inclusive reaction
$a+b\to c+X$ is written
$$\sigma(a+b\to c+X)= \sum_{ijk}F_{i/a}(x_1,Q^2) F_{j/b}(x_2,Q^2)\
\widehat\sigma(ij\to k)\ D_{c/k}(z,Q^2)\eqn\fact$$
for large values $Q$ of the transverse momentum and/or mass of the final
hadron c. The factorization of the cross section into a product of soft
single parton structure functions $(F)$, hadronization functions $(D)$ and
a hard parton-level cross-section $(\widehat\sigma)$, is a remarkable
simplification. The reaction rate depends only on the probability of
finding a single parton in each colliding particle, and the fragmentation
of the final parton is independent of the nature of the target,
projectile and the hard subprocess.

\REF\bh{
S. J. Brodsky and P. Hoyer, \PL B298 1993 165 . See also M. Gyulassy and M.
Pl\"umer, \PL B243 1990 432 ; M. Gyulassy and X. N. Wang, Columbia preprint
CU-TP-598 (June 1993).}

A particular consequence of the leading twist Eq. \fact\ is that the
projectile structure function $F_{i/a}(x_1)$ is independent of the nature of
the target $b$. At first this appears surprising, since one might expect the
projectile parton $i$ to suffer energy loss from gluon radiation while
penetrating a large nuclear target $b$, en route to its hard interaction.
However, the nucleus-induced energy loss is suppressed at high
energies\refmark\bh.

To see the physical reason for the low energy loss, consider a typical Fock
state of the projectile hadron wave function. It consists of the parton that
is going to have the hard collision and a number
of spectator partons. Now at high energies, essentially due to time
dilation, the various partons in the Fock state generally do not have time
to communicate (interact) with one another while in the target --- the Fock
states do not mix. The mixing time (or ``life-time'') of a Fock state is
given by the inverse of the energy difference between the mixing states.
Thus, the emission of a gluon from a quark, $q\to q+G$, involves a kinetic
energy difference $\Delta E$ between the $q$ and the $qG$ states given at
high energy by
$$\Delta E(q\to qG) \simeq {1\over 2 p} \left
[m_q^2-{m_q^2+p_\perp^2 \over 1-x}-{p_\perp^2\over x} \right ] \eqn\ediff$$
where the gluon carries transverse momentum $p_\perp$ and a fraction $x$ of
the initial quark momentum $p$. For large $p$, the emission of a gluon with
moderate $x$ and $p_\perp$ thus requires a long formation time $\sim
1/\Delta E\ \propto p$.

According to the above, typical incoming Fock states (with a small
$\Delta E$ compared to the bound state energy) should be thought of as having
formed long before the target, with a distribution among the Fock states that
depends solely on the projectile wave function. The Fock states do not mix
in the target region (\eg, by emitting or absorbing gluons). Hence the
partons of the Fock state only scatter
elastically while in the nuclear target. In particular, the parton
that is to suffer the hard collision cannot lose energy by emitting gluons of
moderate $p_\perp$ and $x$, if the emission time $1/\Delta E \gsim
R_b$, the radius of the target. It may emit hard (large $p_\perp$) gluons at
the hard collision vertex, but this probability is independent of the
nuclear size and taken into account via higher order terms in the
perturbative expansion of $\widehat\sigma$ in Eq. \fact. The quark may also
emit hard gluons (such that $1/\Delta E \lsim R_b$) somewhere else in the
nucleus. This double parton scattering does depend on the nuclear size, but
is suppressed by the small cross-section for two independent hard
scatterings.

The same argument for limited energy loss applies to the final state. The
partons that emerge from the hard collision have a typical $p_\perp \sim Q$
and hadronize into jets long after leaving the target. At the time of
hadronization, the scattered parton $k$ in Eq. \fact\ is thus far from the
other, ``spectator'' partons in the incoming Fock states, ensuring that the
fragmentation function $D_{c/k}(z)$ is independent of the target and
projectile.

\REF\bbl{G. T. Bodwin, S. J. Brodsky and G. P. Lepage, \PR D39 1989 3287.}

A bound on the parton energy loss which follows from the
above arguments (\ie, the uncertainty principle) is given in
Ref. \bh. It depends on the average hardness $\vev{k_\perp^2}$ of the
collisions that the parton experiences while traversing the nucleus, since
the gluons it radiates have\refmark\bbl\ transverse momenta $p_\perp \leq
k_\perp$. Gluons emitted in the nucleus by a parton of momentum
$p$ can thus carry at most an energy fraction
$$x \lsim {k_\perp^2 R_b\over 2p}\eqn\eloss$$
where $R_b$ is the target radius
($\propto A^{1/3}$ for a target of nuclear number $A$). Assuming
$\vev{k_\perp^2} \sim 0.1\ \gev^2$ one finds that energy loss is negligible
for high energy partons. In fact, even as its momentum $p \to \infty$ the
parton suffering a hard collision can lose only a finite amount of energy in
the nucleus.

\REF\muon{W.  Busza,
\NP A544 1992 49c, proceedings of the Ninth International
Conference on Ultra-Relativistic Nucleus-Nucleus Collisions:
Quark Matter '91,  Gatlinburg, TN (November,1991).}

\REF\ns{N. Schmitz, Munich preprint MPI-PhE/92-23,
talk at the XXII Int. Symp. on Multiparticle Dynamics, Santiago de
Compostela, Spain, July 1992.}

\REF\emc{EMC Collaboration, A.  Arvidson \etal, \NP B246 1984 381 ; J.
Ashman \etal, \ZP C52 1991 1 ; R.  Windmolders, Proc. of the 24.  Int.  Conf.
on High Energy Physics, (Munich 1988, Eds.  R.  Kotthaus and J.  H.  K\"uhn,
Springer 1989), p. 267.}

\REF\dy{D.  M.  Alde \etal, \PRL 64 1990 2479.}

\REF\pa{C. Stewart \etal, \PR D42 1990 1385.}

\REF\bg{A. Bia\l as and M. Gyulassy, \NP B291 1987 793 ; A. Bia\l as and J.
Czyzewski, \PL B222 1989 132.}

Experimentally, there is convincing evidence that the energy loss of high
energy quarks is insignificant (for recent reviews, see
Refs. \muon, \ns). The evidence comes from several different processes: deep
inelastic lepton  scattering$^{\,\emc}$, lepton pair
production$^{\,\dy}$ and large $p_\perp$ jet production in $pA$
collisions$^{\,\pa}$. In deep inelastic lepton scattering, there are signs of
energy loss when the energy $\nu$ transferred to the struck quark in large
nuclei is below about 80 GeV. The $\nu$ dependence of the energy loss has
been studied using models for quark hadronization inside
the nucleus$^{\,\bg}$.

\REF\gm{S.  Gavin and J.  Milana, \PRL 68 1992 1834.}

\REF\qu{E.  Quack, Heidelberg preprint HD-TVP-92-2 (June 1992).}

\REF\ff{S. Frankel and  W. Frati, University of Pennsylvania preprint
UPR-0499T (May 1992).}

\REF\bad{J.  Badier \etal, \ZP C20 1983 101.}

\REF\kat{S.  Katsanevas \etal, \PRL 60 1988 2121.}

\REF\ald{D.  M.  Alde \etal, \PRL 66 1991 2285 ; \PRL 66 1991 133.}

\REF\bho{S. J. Brodsky and P. Hoyer, \PRL 63 1989 1566.}

\REF\bhmt{S.  J.  Brodsky, P.  Hoyer, A.  H.  Mueller and W.-K.  Tang,
\NP B369 1992 519 .}

\REF\vbha{R. Vogt, S. J. Brodsky, and P. Hoyer,
Nucl. Phys. {\bf B360} (1991) 67.}

\REF\ph{P. Hoyer, Acta Phys. Pol. {\bf 23} (1992) 1145.}

Parton energy loss has been invoked$^{\,\gm,\qu,\ff}$ to explain the
suppression of $\Jpsi$ production observed$^{\,\bad,\kat,\ald}$ at large
Feynman $x$ for heavy nuclear targets. The amount of energy loss required
to explain the suppression this way is, however, considerably higher than
allowed by Eq. \eloss. In Refs. \bho, \bhmt\ and \vbha\ a different
explanation for the $\Jpsi$ suppression at large $x$ is proposed, which we
briefly discuss in Sect. 2.4 (see also the review of Ref. \ph).

\vskip 1cm
{\bf 2. Spectator interactions}

In the leading twist approximation the cross section depends only
on {\it single} parton structure functions: The hard process is described
as an incoherent sum of scatterings between one parton in the
projectile and one in the target. There are kinematic situations, however,
where interactions involving more than one parton in the projectile and/or
in the target may be expected, and experimentally are observed, to be
significant. At high energies, this is typically the case when one parton,
or hadon, carries a large fraction $x$ of the energy of its parent. Next I
would like to briefly discuss different aspects of these higher twist
effects.

\REF\bmu{S. J. Brodsky and A. H. Mueller, \PL 206B 1988 285 ;
R. Vogt and S. Gavin, \NP B345 1990 104.}

\REF\hl{R. C. Hwa, \PR D22 1980 1593.}

\vskip .5cm
{\it 2.1 Soft Recombination with Comovers}

At large scales $Q$ of the hard interaction, most partons that are involved
in the hard collision have transverse momenta of $\O(Q)$. Hence they do not
interact significantly with the spectator partons, which have limited
transverse momenta. However, a small fraction of the ``hard'' partons, of
$\O(\Lambda_{QCD}^2/Q^2)$, will be emitted with limited transverse
momenta. They can then interact strongly with comoving spectators of
similar velocities$^{\,\bmu}$, and in particular coalesce with
them to form hadrons. A specific model including such parton recombination
has been studied already several years ago\refmark\hl.

Parton coalescence has two principal effects:
\item{(i)} No momentum is lost in the hadronization -- the hadron momentum
is the sum of the momenta of the coalescing partons. Hence the $x_F$
distribution is harder than that described by the decay function $D(z)$ in
Eq. \fact.
\item{(ii)} There will be quantum number correlations (``leading particle
effects'') between the projectile and the produced hadron.

Note that the coalescence effects are expected to be relevant only for those
hadrons that are procuced with a limited transverse momentum. Hence the
hadron $p_\perp$ distribution should steepen with $x_F$, and only the
low $p_\perp$ hadrons should show the quantum number correlations.

\REF\bia{S. F. Biagi, \etal, {Z.~Phys.~{\bf C28}\ (1985) 175}.}
\REF\cha{P. Chauvat, \etal, {Phys.~Lett.~{\bf 199B}\ (1987) 304}.}
\REF\cot{P. Coteus, \etal, {Phys.~Rev.~Lett.~{\bf 59}\ (1987) 1530}.}
\REF\shi{C. Shipbaugh, \etal, {Phys.~Rev.~Lett.~{\bf 60}\ (1988) 2117}.}
\REF\agu{M. Aguilar-Benitez, \etal, {Z.~Phys.~{\bf C40}\ (1988) 321}.}

\REF\app{J. A. Appel, FERMILAB-Pub-92/49, to appear in Ann. Rev. Nucl.
Part. Sci. {\bf 42} (1992);
G. A. Alves, \etal, FERMILAB-Pub-92/208-E (August, 1992).}
\REF\wa{M. Adamovich \etal, WA82 Collaboration, Contribution to ICHEP 92
(Dallas, Texas, August 1992).}

\REF\chr{S. Bethke, {Z.~Phys.~{\bf C29}\ (1985) 175};
JADE Collaboration, {Z.~Phys.~{\bf C33}\ (1987) 339};
J. Chrin, {Z.~Phys.~{\bf C36}\ (1987) 163};
D. Decamp, \etal, {Phys.~Lett.~{\bf 244B}\ (1990) 551}.}

\REF\vbhb{R. Vogt, S.J. Brodsky and P. Hoyer, \NP B383 1992 643.}

Evidence for comover effects have been seen, in particular, in charm
hadroproduction. Since the charm quark mass is not very large, a considerable
part of the charm quarks will be produced with transverse momenta small
enough for their hadronization to be affected by comoving spectator quarks.
Strongly enhanced charm production at large $x_F$ has been reported by
several experiments$^{\,\bia,\cha,\cot,\shi,\agu}$, but the low
statistics and unconfirmed status of the observations long prevented
definite conclusions. More recently, data$^{\,\app,\wa}$ on $\pi^-A \to D+X$
established that the $x$ distribution of the $D$ meson has the same shape as
the charm $quark$ distribution predicted by the leading twist Eq. \fact.
Hence the decay function $D(c \to D)$ must be assumed to be approximately
$\delta(1-z)$ in hadroproduction, significantly different from that
measured$^{\,\chr}$ in $e^+e^- \to D +X$. This agrees with the comover
effect$^{\,\vbhb}$, since the coalescence of a charm quark with a light
quark of similar velocity gives a $D$ meson of momentum similar or slightly
higher than the charm quark. The data$^{\,\app,\wa}$ is also consistent with
the model of Ref. \vbhb\ for quantum number correlations between the
projectile and produced $D$ meson.

It should be stressed that comover coalescence is a soft
process that is not calculable in perturbative QCD. Due to the softness of
the interaction, the recombination cannot significantly change the
momentum of a heavy quark. Processes where, \eg, a fast light quark
combines with a slow heavy quark, thus giving rise to a fast heavy hadron,
require large momentum transfers and $should$ be calculated using PQCD. We
shall return to this question below.

\REF\leit{D. M. Alde, \etal, Phys.~Rev.~Lett. {\bf 66} (1991) 2285; M. J.
Leitch, \etal, Nucl.~Phys.~{\bf A544} (1992) 197c.}

Quarkonia contain no light quarks, and thus are not
formed by soft coalescence with valence quarks. On the other hand, the
formation of quarkonia at moderate $x_F$ is expected to be suppressed by
comovers interactions, which can break up the fragile bound state, resulting
in open heavy flavor production$^{\,\bmu,\vbhb,\vbha}$. Evidence for this
has been seen$^{\,\leit}$ in $\Jpsi$ and $\Upsilon$ prodution on
nuclear targets, in the nuclear fragmentation region $(x_F \lsim 0)$. The
suppression of quarkonium production is observed to increase with nuclear
number, which is consistent with the increasing number of comoving
spectators.

\vskip .5cm
{\it 2.2 Hard Higher Twist Processes}

The probability of finding two partons within a short transverse distance
$1/Q$ in typical Fock states is for geometrical reasons of order $1/Q^2$.
Since $1/Q$ is the coherence length of hard processes, this accounts for
the size of the higher twist corrections to the single parton scattering
Eq. \fact. However, there are processes which only get contributions from
transversally compact Fock states. In this case the corrections to the
leading twist formalism can be substantial, and even dominant. A simple
example is deep inelastic lepton scattering at large Bjorken $x$. A Fock
state in which one quark carries $x \simeq 1$ has, according to Eq. \ediff,
a large energy and hence mixes rapidly with other Fock components.
Intuitively, the partons that transferred their momentum to the leading
quark must have been nearby, to accomplish the transfer in the short
life-time of the $x \simeq 1$ state.

\REF\bhmt{S. J. Brodsky, P. Hoyer, A. H. Mueller, W.-K.Tang, Nucl. Phys.
{\bf B369} (1992) 519.}

More quantitatively$^{\,\bhmt,\ph}$, consider a parton that is going to
transfer its longitudinal momentum, say $yp$, to the leading quark from
which the lepton then scatters. This parton can have any transverse momentum
up to a large value $k_\perp$ for which the lifetime of its Fock state,
$2yp/(m^2+k_\perp^2)$, becomes similar to the lifetime of the $x \simeq 1$
state, $2p(1-x)/(m^2+p_\perp^2)$. For finite $y$ and limited transverse
momentum $p_\perp$ of the $x \simeq 1$ parton, this shows that $k_\perp^2
\propto 1/(1-x)$. Hence the transverse size of the Fock states in the target
hadron which can contribute to a deep inelastic scattering event with $x
\simeq 1$ scales as $r_\perp \propto \sqrt{1-x}$.

Since the Fock states that participate in large $x$ deep inelastic
lepton scattering have a small transverse size, the higher twist
corrections due to coherent scattering from more than one quark in
the target hadron can be important. In particular, in the limit where the
coherence length $1/Q$ scales as the transverse size $r_\perp \sim
\sqrt{1-x}$, the twist expansion breaks down. Higher twist contributions are
not suppressed at all compared to leading twist for arbitrarily large
$Q^2$, if $x \to 1$ as $Q \to \infty$ such that $\mu^2 = Q^2(1-x)$ is held
fixed$^{\,\bhmt}$.

\REF\vm{M. Virchaux and A. Milsztajn, Phys.~Lett.~{\bf B274}\ (1992) 221.}

\REF\exc{S. J. Brodsky and G. P. Lepage, published in {\it Perturbative
QCD}, edited by A.H. Mueller, World Scientific, (1989).}

Experimentally, the rise of the higher twist terms with increasing
Bjorken $x$ is clearly seen in deep inelastic scattering\refmark\vm.
Since the hard scattering from the transversally small target can be
calculated in PQCD, the data can, in principle, be used to obtain information
on multiparton correlations in the proton. The large $x$ cross
section measures\refmark\bhmt\ the target distribution function -- the
probability to find all valence partons at small transvese separations,
which is a function of their sharing of the total longitudinal momentum. The
same distribution function also determines hard exclusive
cross-sections\refmark\exc, \eg, elastic lepton hadron scattering at large
$Q^2$.

\REF\nz{N. N. Nikolaev and B. G. Zakharov, \ZP C49 1991 607.}
\REF\dbh{V. Del Duca, S. J. Brodsky and P. Hoyer, \PR D46 1992 931.}
\REF\iof{B. L. Ioffe, \PL 30B 1969 123.}

\vskip .5cm
{\it 2.3 Photoproduction of hadrons at large $x_F$}

The above Fock state picture is useful also for understanding the
interactions of photons. In deep inelastic scattering, the splitting of a
photon of energy $\nu$ and virtuality $-Q^2$  into a $q \bar q$ pair involves
an energy difference
$$\Delta E = -Q^2 - {1\over 2\nu}\, {m_q^2+p_\perp^2 \over
z(1-z)}\eqn\photon$$
For large $Q^2$ and $\nu$, $\Delta E$ is essentially independent of the
fraction $z=E_{\bar q}/\nu$ of the photon energy carried by the antiquark.
Since a slow antiquark interacts strongly in the target, the DIS
cross-section is dominated\refmark{\nz,\dbh} by $z \sim \O(1/\nu)$. This
picture of deep inelastic scattering in the target rest frame also gives
an understanding of the nuclear shadowing at small $x_{Bj}=Q^2/2M\nu$, since
the transverse size of the Fock state turns out to grow like $1/x_{Bj}$,
the ``Ioffe'' distance\refmark\iof\ from the target at which the photon
creates the $q\bar q$ pair. In particular, one finds\refmark{\nz,\dbh} that
the amount of shadowing depends on the polarization of the virtual photon.
This prediction has not yet been tested experimentally.

For real photoproduction, $Q^2=0$, an equal partition of the energy $(z
\simeq \half)$ minimizes the energy difference $\Delta E$ in Eq. \photon.
These $q\bar q$ states are longlived and develop into meson states before
the collision --  hence the photon may be represented as a mixture of vector
mesons according to the Vector Dominance Model.

\REF\bart{D. S. Barton \etal, \PR D27 1983 2580.}

The pointlike photon manifests itself, however, when special requirements
are imposed on the final state. Hard reactions involving the production of
heavy quarks or large transverse momenta are the most well-known examples.
Another, less obvious case\refmark{\muon,\bh} is that of photoproduction
of inclusive hadrons with limited $p_\perp$ but large Feynman
$x_F=E_h/\nu$.  As discussed above, this process also involves a short time
scale. The photon can most easily create a parton with large momentum via
an asymmetric $q \bar q$ decay, $z \propto 1-x_F$ in Eq. \photon.
The quark (or antiquark) with large momentum will then penetrate the nucleus
with minimal energy loss, as discussed in Section 1. This means that the
large $x_F$ distribution of photoproduced hadrons will be independent of the
nuclear size, contrary to the case for hadronic projectiles where an $x_F$
dependent nuclear suppression is observed\refmark\bart. The data on
muon scattering shows\refmark\muon\ that the inclusive hadron
$x_F$ distribution has the same shape for all nuclear targets, and is
furthermore independent of $x_{Bj}$ and $Q^2$. In particular, the shape of
the $x_F$ distribution is the same in and out of the shadowing region.
This agrees with the present picture\refmark\bh, according to which the
pointlike photon should manifest itself even at $Q^2=0$ when $x_F$ is large.

\vskip .5cm
{\it 2.4 Heavy quark production at large $x_F$}

In the standard, leading twist picture of heavy quark production
represented by Eq. \fact, the heavy quarks obtain their momenta from a
single parton in each of the colliding hadrons. As we saw in Section 2.2, at
large $x_F$ only transversally small Fock states of the projectile
contribute. When the transverse size $r_\perp\propto \sqrt{1-x}$ of the Fock
state is commensurate with the Compton wavelength $1/M$ of the heavy quark,
``intrinsic'' diagrams, where the heavy quark pair is connected to several
partons in the projectile, become important\refmark{\bhmt,\ph}. Furthermore,
the target parton scatters mainly from the stopped, light valence
partons (which have momenta of $\O(1-x_F)$. The hardness of the interaction
with the target is $\mu^2 = M^2(1-x_F)$, which can be small even for large
quark masses $M$.

\REF\cip{C. Biino \etal, \PRL 58 1987 2523.}
\REF\fal{S. Falciano \etal, \ZP C31 1986 513.}
\REF\hei{J. G. Heinrich \etal, \PR D44 1991 1909.}
\REF\ber{E. L. Berger and S. J. Brodsky, \PRL 42 1979 940.}

So far no complete QCD calculation has been done for heavy quark production
in the limit of fixed $\mu^2$. General aspects of the large $x_F$
$\Jpsi$ production\refmark{\bad,\kat,\ald,\cip}, as well as model
calculations\refmark\vbha, nevertheless suggest that this limit is relevant
for understanding the data. In particular, the nuclear target
$A$ dependence, the decrease of the average transverse
momentum of the $\Jpsi$ and the Feynman scaling of the cross section all
support this view\refmark{\vbhb,\vbha,\bhmt,\bho}, and disagree with the
leading twist prediction. Furthermore, the fact that the produced $\Jpsi$ is
unpolarized except for $x_F \gsim 0.9$, where it is longitudinally
polarized\refmark\cip, suggests a different production mechanism at large
$x_F$. An analogous change of polarization is seen in large mass lepton pair
production\refmark{\fal,\hei} at high $x_F$, and is understood as
due to higher twist effects\refmark\ber.

\REF\biab{S. F. Biagi \etal, \ZP C34 1987 187.}

A particularly striking example of spectator effects is provided by a
comparison of the inclusive production of $\Omega^-$ in $pN$ and $\Xi N$
collisions\refmark\biab. The inclusive $pN \to\Omega^-+X$ cross-section falls
steeply with $x_F$, as would be expected. On the contrary, the $\Xi N
\to\Omega^-+X$ cross-section $rises$ by more
than two orders of magnitude to a maximum at $x_F \sim 0.8$. At its maximum,
the $\Xi N \to\Omega+X$ cross-section is comparable to that of $pN \to
\Lambda X$, and more than four orders of magnitude larger than the $pN
\to \Omega$ corss-section. Clearly the ``spectator'' strange
quarks in the $\Xi$ projectile are combining with a produced strange quark
to form the $\Omega^-$.

The strange quark is too light to trust the hard scattering formalism in a
quantitative way. Nevertheless, the behavior of the $\Xi N
\to\Omega+X$ process agrees qualitatively with what one would expect if the
strange quark were truly massive. In that case, the two strange quarks in the
projectile would each carry one half of the incident $\Xi$ momentum, while
the three strange quarks in the $\Omega^-$ would similarly each carry
$x_F/3$. Thus the incident strange quarks need to be decelerated, which
requires large momentum transfers if they are considered to be heavy.
A maximal $\Omega^-$ inclusive cross-section is expected when the
deceleration is as small as possible, which implies $x_F \simeq .75 \ldots
1$, depending on the momentum of the $\bar s$ quark that is also produced in
the collision.

\vskip 1cm
{\bf Acknowledgements}

The material presented here is the result of collaborations and discussions
with, in particular, Stan Brodsky, Vittorio Del Duca, Al Mueller, Wai-Keung
Tang, Ramona Vogt and Mikko V\"anttinen. I am also grateful to the
organizers of the Cracow Workshop for their invitation and warm hospitality.

\refout
\bye